\documentclass[twocolumn,english,showpacs]{revtex4}
\usepackage{babel,amsmath,amssymb,dcolumn}
\usepackage[dvips]{graphics}

\begin{document}

\title{Magnetised black hole as a gravitational lens}

\author{R.A. Konoplya}
\email{konoplya@fma.if.usp.br}
\affiliation{Instituto de F\'{\i}sica, Universidade de S\~{a}o Paulo \\
C.P. 66318, 05315-970, S\~{a}o Paulo-SP, Brazil}


\begin{abstract}
We use the Ernst-Schwarzschild solution for a black hole immersed
in a uniform magnetic field to estimate corrections to the bending
angle and time delay due-to presence of weak magnetic fields in
galaxies and between galaxies, and also due-to influence of strong
magnetic field near supermassive black holes. The magnetic field
creates a kind of confinement in space, that leads to increasing
of the bending angle and time delay for a ray of light propagating
in the equatorial plane.
\end{abstract}

\maketitle




\section{1. Introduction.}

One of the most intriguing observable test of general relativity
is deflection of light exerting gravitational attraction. This
effect is feasible for observation nowadays in many scales: deflection
induced by compact object, such as stars or black holes, by
galaxies, cluster of galaxies, and even by the large scale
structure of the universe \cite{Lens}. At the same time, an important factor
which influences the dynamic of the interacting matter in the
Universe is existing of magnetic fields: from weak fields ($\sim 5
\mu G$) in galaxies, cluster of galaxies ($\sim 6
\mu G$) and between clusters of galaxies ($\sim 50
n G$), until very strong magnetic fields near super-massive black
holes. This fields should influence not only propagation of
charged matter, but also of rays of light at large distances or in
the regime of strong gravitational field, when the magnetic field
is one of the parameters of the space-time metric.

In this letter, we shall study the corrections to the deflection angle
and time delay for a ray of light in vicinity of black holes
immersed in a magnetic field. The only known exact solution of the
Einstein-Maxwell equations describing the non-rotating black hole
immersed in a magnetic field is the Ernst solution \cite{Ernst}.
Thermodynamical and geometrical properties of this solution were
investigated in \cite{Ernst-2}. Some interesting generalisations of the
solution were found in \cite{Ernst-gen}, and different effects near
magnetised black hole were studied in \cite{Aliev}.

We shall use Ernst solution as a model for our estimations of the
corrections. The problem is solvable in the equatorial plane where
the equations of motion allow separation of variables. The main
feature we observed is that the magnetic field, even being small
as a space-time metric parameter, i.e. small enough not to deform
the geometry of the compact objects,  leads to considerable
increasing of deflection angle and time delay when distance from
an observer to a source is large enough. The strong magnetic field
localised near black holes also induces increasing of deflection
angle and time delay.

\section{Null geodesics around Ernst-Schwarzschild black holes}

The exact solution of the Einstein-Maxwell equations, represented
by the Ernst generalisation for a Schwarzschild metric
\cite{Ernst}, describes a non-rotating black hole immersed in an
asymptotically homogeneous external magnetic field, and the
corresponding metric has the form:
\cite{Ernst}:
$$
d s^{2} = \Lambda^{2} \left(
\left(1- \frac{2 M}{r} \right) d t^{2} - \left(1- \frac{2 M}{r} \right)^{-1} d
r^{2} -r^{2} d \theta^{2}
\right)
$$
\begin{equation}
 - \frac{r^{2} \sin^{2} \theta}{\Lambda^{2} } d \phi^{2}
\end{equation}
where the external magnetic field is determined by the real
parameter $B$, and
\begin{equation}
\Lambda = 1 + B^{2} r^{2} \sin^{2} \theta.
\end{equation}
The vector potential for the magnetic field is given by the
formula:
\begin{equation}
A_{\mu} d x ^{\mu} = - \frac{B r^{2} \sin^{2} \theta} {2 \Lambda}
d \phi.
\end{equation}

As a magnetic field is assumed to exist everywhere in space, the
above metric is not asymptotically flat. The event horizon is
again $r_{h} = 2 M$, and the surface gravity at the event horizon
is the same as that for a Schwarzschild metric, namely
\begin{equation}
\chi = 2 \pi T_{H} = \frac{1}{4 M}.
\end{equation}
This leads to the same thermodynamic properties \cite{Radu} as for the
case of Schwarzschild black hole.

To analyse the lens parameters of the above black hole, we need to
consider the null geodesic equations for Ernst-Schwarzschild
space-time \cite{Galtsov-Ernst}. For this, let us introduce the
momenta:
\begin{equation}
p_{\mu} = g_{\mu \nu} \frac{d x^{\mu}}{d s},
\end{equation}
where $s$ is an invariant affine parameter. The Hamiltonian is
\begin{equation}
H=\frac{1}{2} g^{\mu \nu} p_{\mu} p_{\nu}.
\end{equation}
The action can be represented in the form:
\begin{equation}
S = E t - L \phi - S_{r}(r) - S_{\theta}(\theta),
\end{equation}
where $E$ and $L$ are the particle`s energy and angular momentum
respectively.

Then, the Hamilton-Jacoby equations for null geodesics read
\begin{equation}
\frac{1}{2} g^{\mu \nu} \frac{\partial S}{\partial x^{\mu}} \frac{\partial S}{\partial
x^{\nu}} = - \frac{ \partial S} {\partial s} = 0.
\end{equation}

It is evident that the equations of motions allow separation of
variables in the equatorial plane $\theta = \pi/2$
\cite{Galtsov-Ernst}. After making use of the equation
$ p^{1} = - p_{0}/g_{00} (d r/d t)$ and equation (8), we easily
find the propagation equation
\begin{equation}
\left(\frac{d r}{d t}\right)^{2} = -\frac{g_{00}}{g_{11}} \left(1 + \frac{g_{00}}{g_{33}}
\left(\frac{p_{3}}{p_{0}}\right)^{2} \right),
\end{equation}
where $p_{0} = E$, and $p_{3} = - L$ are constants of motion, and
$g_{i k}$ are metric coefficients given by formulas (1) and (2) at $\theta =\pi/2$. In a
similar fashion we find the geodesic trajectory equation
\begin{equation}
\left(\frac{d r}{d \phi}\right)^{2} = -\frac{g_{33}}{g_{11}} \left(1 + \frac{g_{33}}{g_{00}}
\left(\frac{p_{0}}{p_{3}}\right)^{2} \right).
\end{equation}
We see that propagation and trajectory equations contain only
ratio $b = L/E$, which is called the impact parameter. The
qualitative description of the motion can be made by considering
the effective potential of the motion:
\begin{equation}
U_{eff} = \pm \frac{g_{00}L^{2}}{g_{33}}.
\end{equation}
The solution of the equations $U_{eff}=\partial U_{eff}/ \partial
r = 0$ gives the radius the closed circular geodesic orbits. The
essential feature of this potential for Ernst-Schwarzschild case
is that for non-zero $b$, the escape of the massless particles in the
equatorial plane is impossible. This is stipulated by existence of an additional
turning point (where $dr/dt = 0$). For small magnetic fields, we
will be interested here, this additional turning point occurs at a
sufficiently large $r$.

\vspace{3mm}
\section{Bending angle and time delay for Ernst-Schwarzschild black holes}

Passing near black hole ray of light approaches the black hole at
some minimal distance from it, called distance of closest approach
$r_{min}$. In Schwarzschild  case, $r_{min}$ is determined as the
largest real root of the equation $dr/dt = 0$. Yet, in presence of
magnetic fields, $r_{min}$ is not the largest root anymore: In the
regime of small $B$, the largest root corresponds to another
turning point very far from black hole. Thus, $dr/dt = 0$ gives
\begin{equation}
r^3 + (2 M - r) (1 + B^{2} r^{2})^{4} b^{2} = 0
\end{equation}

If we know $r_{min}$ with great accuracy, we can perform
integrations for finding bending angle:
\begin{equation}
\alpha = \phi_{s}-\phi_{o} = -\int_{r_{s}}^{r_{min}} \frac{d \phi}{d r} dr +
\int_{r_{min}}^{r_{o}} \frac{d \phi}{d r} dr - \pi.
\end{equation}
Here $r_{o}$ is radial coordinate of an observer and  $r_{s}$ is
radial coordinate of the source.

In a similar fashion one can find the time delay, which is the
difference between the light travel time for the actual ray, and
the travel time for the ray the light would have taken in the
Minkowskian space-time:
\begin{equation}
t_{s}-t_{o} = -\int_{r_{s}}^{r_{min}} \frac{d t}{d r} dr +
\int_{r_{min}}^{r_{o}} \frac{d t}{d r} dr - \frac{d_{s-o}}{\cos \mathcal{B}} .
\end{equation}
Here the term  $\frac{d_{s-o}}{\cos \mathcal{B}}$ represents the
propagation time for a ray of light, if the black hole is absent
(see for instance \cite{keeton1}).

We shall also use the following designation: the difference in
time delay between Schwarzschild and Ernst - Schwarzschild space-times
\begin{equation}
\delta = (t_{s}-t_{o})_{E-S} - (t_{s}-t_{o})_{S}.
\end{equation}

Equations (9) and (10) can be re-written in the following form:
\begin{equation}
 \left(\frac{d t}{d r}\right) = \frac{\left(1- \frac{2 M}{r}\right)^{-1}}{\sqrt{1 +  \frac{(2 M - r)(1+ B^{2}r^{2})^{4} b^{2}}{r^{3}}}}
\end{equation}
\begin{equation}
\left(\frac{d \phi}{d r}\right) = \frac{(1+ B^{2}r^{2})^{4}}{\sqrt{(2 M - r) r (1+ B^{2}r^{2})^{4}
+ \frac{r^{4}}{b^{2}}}}
\end{equation}

Now, using (16), (17) in integrals (13) and (14), we are in
position to compute the bending angles and time delays for
different values of the impact parameter $b$ and magnetic field $B$.
First, the parameter $B$ of the external magnetic field in (1-2),
and magnetic field in units of Gauss are connected by the
following relation:
\begin{equation}
B = 4.25 \cdot 10^{-21} \frac{B_{0}}{M_{\bigodot}},
\end{equation}
where $B_{0}$ is the the magnetic field in units of Gauss. As $B$
has units of the inverse length, sometimes natural units are used
$B = 1/r_{h}$, were $r_{h} = 2 M$. A super-massive black hole in
the centre of our galaxy has the mass $M = (3.6 \pm 0.2) 10^{6}
M_{\bigodot}$ (what corresponds to $M
\approx 3.4 \cdot 10^{-7}$ pc, in units c=G=1), and is situated at
a distance $r_{o}= (7.9 \pm 0.4) 10^{3}$ pc from the Earth
\cite{BHdistance}. The "unit" magnetic field $B = 1/(2 M)$ for the
above super-massive black hole is $B_{0} = 3.268 10^{13} G$,
according to the formula (18).

Typical distance from a source to a black hole is of order $r_{s}
= 1$ pc. Now let us take $M = 1$, then we can re-scale the
corresponding values for $r_{o}$ and $r_{s}$
\begin{equation}
r_{o}= 2.3235 \cdot 10^{10}, \quad r_{s} = 2.9412 \cdot 10^{6},
\quad M=1.
\end{equation}
\vspace{3mm}

\begin{center}
\begin{table*}
\caption{Bending angle $\alpha + \pi$ and value of the "far" turning point $r_{tun}$,
evaluated for different values of the impact parameter $b$ and
small magnetic field $B$ for the super-massive black hole in the
centre of our galaxy, $r_{o}= 2.3235 \cdot 10^{10}$, $r_{s} =
2.9412
\cdot 10^{6}$. (Geometrical units are used, $M=1$).}
\begin{tabular}{|c|c|c|c|c|c|c|}
  \hline
  $B$ & $\alpha + \pi$ ($b=6$) & $r_{tun} \cdot 10^{12}$  & $\alpha + \pi$ ($b=10$) & $r_{tun} \cdot 10^{12}$
  & $\alpha + \pi$ ($b=20$) & $r_{tun} \cdot 10^{12}$ \\
  \hline
  $ 0 \cdot 10^{-10}$ &  4.8609788632 & --& 3.7319849720 & -- & 3.37772180250 & --  \\
  $ 1 \cdot 10^{-10}$ &  4.8609789477 & 11.8563 & 3.7319851128 & 9.99999 & 3.37772208405 & 7.93700  \\
  $ 2 \cdot 10^{-10}$ &  4.8609892297 & 4.70518 & 3.7320022496 & 3.96850 & 3.37775635750 & 3.14980  \\
  $ 3 \cdot 10^{-10}$ &  4.8612094483 & 2.74023 & 3.7323692805 & 2.31120 & 3.37849042386 & 1.83440  \\
  $ 4 \cdot 10^{-10}$ &  4.8631698141 & 1.86725 & 3.7356365568 & 1.57490 & 3.38502497645 & 1.25000  \\
  $ 4. 5 \cdot 10^{-10}$ & 4.8665248390 & 1.59588 & 3.7412282650 & 1.34601 & 3.39620839297 & 1.06833  \\
  $ 5 \cdot 10^{-10}$ &  4.8737382850 & 1.38672 & 3.7532506749 & 1.16961 & 3.42025321276 & 0.92832  \\
  $ 5.5 \cdot 10^{-10}$ & 4.8881343306 & 1.22123 & 3.7772440844 & 1.03003 & 3.46824003166 & 0.81753  \\
  $ 6 \cdot 10^{-10}$ &  4.9151533653 & 1.08746 & 3.8222758088 & 0.91720 & 3.55830348461 & 0.727981  \\
\hline
\end{tabular}
\end{table*}
\end{center}

For Ernst model to be valid we need a small enough value of $B$,
so that the {\it distant turning point}, responsible for
"reflecting" of a massless particle from infinity, would be at
least a few orders larger than the largest of $r_{o}$ and $r_{s}$.
Thereby we imply that region of asymptotic behaviour of the
magnetic field is far enough, not to create confinement of light
rays. Indeed, for instance for $B
\sim 10^{-10}$, that turning point occur at $r = r_{tun} \sim 10^{14}$, i.e.
four orders greater than $r_{o}$. In fact the applicability of the
Ernst solution to real situations depends on two factors: the
value of the magnetic field $B$, and the distance at which the
Ernst space-time is considered: if $B$ is large or even of order
$1/M$ this strong magnetic field everywhere in space creates large
potential barrier around black hole. This induces strong
confinement even at a distance not far from black hole. This
situation should be remedied by matching the homogeneous magnetic
field with decreasing magnetic field at some distance from black hole.
At the same time, near the event horizon Ernst solution should be
adequate. When $B$ is much less than $1/M$, we can apply Ernst
model even for sufficiently large distances, when confining
properties of the effective potential show themselves at even
larger scale of distances. When $B \rightarrow 0$, we are
approaching the Schwarzschild limit and can consider larger and
larger distances within Ernst solution as a model for magnetic
field in our Universe.


\begin{figure*}
{\includegraphics{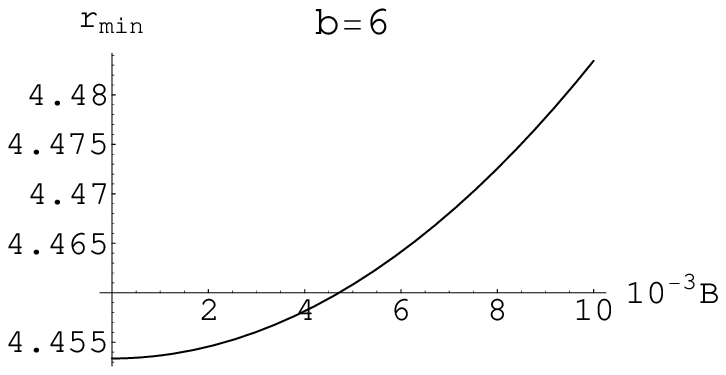}}{\includegraphics{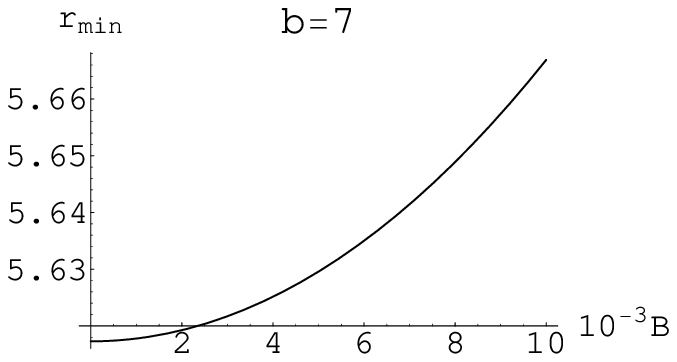}}
{\includegraphics{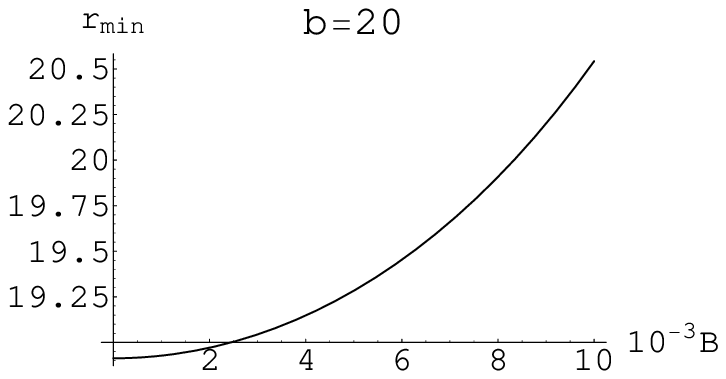}}{\includegraphics{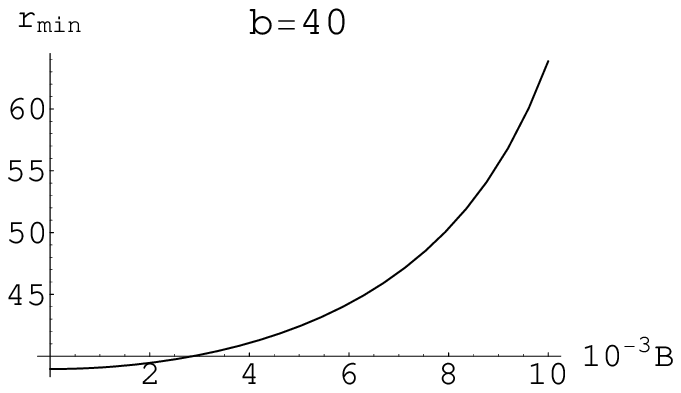}}
\caption{Radius of minimal approach $r_{min}$ as a function of the impact magnetic field strength $B$
for the impact parameters $b = 6$, $b = 7$, $b = 20$, and $b=40$.}
\end{figure*}



\begin{figure*}
{\includegraphics{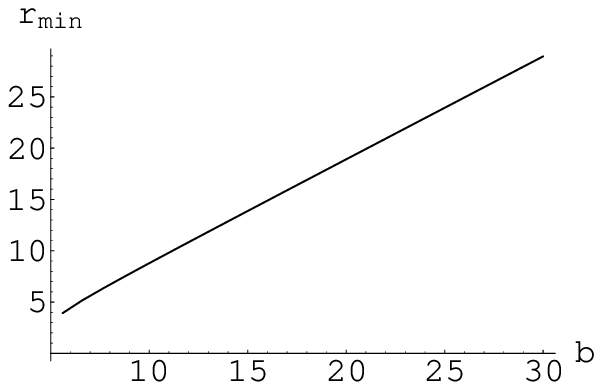}}{\includegraphics{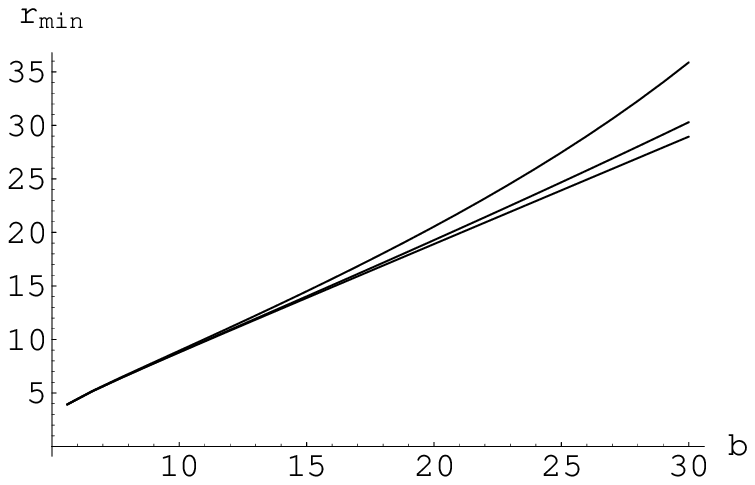}}
\caption{(Left) Radius of minimal approach $r_{min}$ as a function of
the impact parameter $b$ for $B = 10^{-6}$.
(Right). Radius of minimal approach $r_{min}$ as a function of the impact parameter $b$ for $B = 10^{-6}$ (bottom),
$B = 5 \cdot 10^{-3}$, and $B = 10^{-2}$ (top). We see that for
relatively small $B$, $r_{min}$ does not change seemingly when
changing $B$, unless the impact parameter $b$ is large enough.}
\end{figure*}

First, let us consider weak magnetic fields for super-massive black
holes. The results of numerical computation of the integrals (13) and
(14) are given in tables I and II. We integrated expressions (13)
and (14) directly without changing variable using built-in
functions in Mathematica. The integration up to very large values
of limits of integrals is stable, provided one controls the precision
of all incoming data and intermediate procedures. To check this we
can integrate, for instance, (13) for $r_{o}=r_{s}= 10^{15}$
($M=1$), and some large $b$: For Schwarzschild limit the results
show excellent agreement with formula (24) of
\cite{keeton1} in PPN-expansion approach. Thus, for $b=20$ from (13)
we get, $\alpha + \pi = 3.777285$, while (24) in \cite{keeton1}
gives $\alpha + \pi = 3.777144$. This also shows that weak field
limit expanded in higher order in $M/b$ is indeed very good
approximation.

\begin{table}
\caption{Difference in time delay $\delta$ between Schwarzschild and Ernst-Schwarzschild
space-times (in geometrical units, $M=1$) for  small magnetic field $B$
for the super-massive black hole in the centre of our galaxy,
$r_{o}= 2.3235 \cdot 10^{10}$, $r_{s} = 2.9412
\cdot 10^{6}$, $b=6$.}
\begin{tabular}{|c|c|}
  \hline
  $B $ & $ \delta $ ($b=6$) \\
  \hline
 $ 1 \cdot 10^{-10}$&  0  \\
 $ 2 \cdot 10^{-10}$&  0.000031   \\
 $ 3 \cdot 10^{-10}$&  0.000690    \\
 $ 4 \cdot 10^{-10}$&  0.006573  \\
 $ 5 \cdot 10^{-10}$&  0.038277 \\
 $ 6 \cdot 10^{-10}$&  0.162521  \\
 $ 7 \cdot 10^{-10}$&  0.553574  \\
\hline
\end{tabular}
\end{table}

Now let us try to estimate the effect of the strong magnetic field
in the central part of the galactic black hole. For this one cannot
integrate equations (13), (14) up to some large distances, because
as is well-known, the magnetic field near black hole decreases
quickly as the distance is increased. Nevertheless we are able to
consider the bending of light ray and delay in propagation time
{\it near} black hole, implying that {\it far} from black hole the magnetic
field influences the situation according to the estimation scheme
described before.  For this we put an ``observer'' and a ``source''
not far from black hole: on illustrative example in Table III
$r_{o}=r_{s}=20$ ($M=1$).

\begin{table}
\caption{Bending angle $\alpha$ and propagation time $\tau$ for
Ernst-Schwarzschild space-times (in geometrical units, $M=1$)
for $b=6$. "Observer" and "source"  are supposed to be situated
not far from the black hole in order to estimate influence of
magnetic field in the central region of the galactic black hole:
$r_{o} = r_{s}= 20$, $b=6$.}
\begin{tabular}{|c|c|c|}
  \hline
  $B $ & $ \alpha + \pi$ ($b=6$) &  $t_{s} - t_{o} + d_{s-o}/\cos \mathcal{B}$ ($b=6$) \\
  \hline
 $ 0 $              &  4.25233372 &  56.845537725 \\
 $ 10^{-6}$ &    4.25233383 &        56.845537425 \\
 $ 10^{-5}$&     4.25233401  &       56.845538048  \\
 $ 10^{-4}$&     4.25234254  &       56.845559125 \\
 $ 10^{-3}$&     4.25320549  &       56.847699341 \\
 $ 2 \cdot 10^{-3}$&   4.25582382 &  56.854196524 \\
 $ 3 \cdot 10^{-3}$&   4.26019947  & 56.865066714 \\
 $ 4 \cdot 10^{-3}$&   4.26635039  & 56.880373524 \\
 $ 5 \cdot 10^{-3}$&   4.27430179  & 56.900206814 \\
\hline
\end{tabular}
\end{table}

Let us come back in our estimation from geometrical to ordinary units.
We have used Ernst solution for estimation of the influence of the
magnetic fields in {\it two regimes}: {\it weak} galactic magnetic field
existing everywhere in the galaxy and {\it strong} magnetic field in
the region close to the supermassive black hole.

Weak magnetic field is supposed to influence the propagation of
light during the whole way from a source to an observer, and
therefore is expected to make considerable correction to bending
angle and time delay. From table I we see that the increasing in
bending angle due to magnetic field for a case of supermassive
black hole can be considered as observable at the present time
only if the magnetic field would be at least $B \sim 10^{-10}$ or
greater, i.e. about $ \sim 10^{2} G $. This is much larger than
the real galactic magnetic field, which is $\sim 10 \mu G$. The
time delay is affected by weak magnetic field non-negligibly, also
only for $ B_{0} \sim 10^{2} G $ or larger. Yet, even seeming
corrections to time delay and bending angles for $ B_{0} \sim
10^{2} G $ is quite surprising, because such  a field is too weak
to deform the geometry of the Schwarzschild black hole: the
space-time near Ernst black hole with $ B_{0} \sim 10^{2} G $ is,
in fact, a Schwarzschild space-time with very high accuracy. The
magnetic field should be much larger, about $\sim 1/M$ in
geometric units, to create considerable effect near black hole.
Therefore we can conclude that the correction effect at $ B_{0}
\sim 10^{2} G $ is {\it due-to the existence of the non-vanishing
magnetic field everywhere} in the way of a ray of light, and not
because of peculiarities of black hole geometry near the event
horizon.

For the intergalactic magnetic field $\sim 6 \mu G$ (what is $\sim
\cdot 10^{20}$, when $M=1$ and $G=c=1$), the distant turning point
is situated at $r_{tun} = 3.18 \cdot 10^{18} pc $. Therefore we
can expect considerable correction to the bending angle and time
delay only if $r_{o}$ is approaching by order to the radius of the
distant turning point, i.e. at least about $\sim 10^{16}-10^{17}
pc $, what is much greater than the size of our cluster of
galaxies ($\sim 1.5 \cdot 10^{7}pc$).

On the contrary, strong magnetic field in the centre of galaxy decay
quickly with distance and does not make influence far from black hole,
yet it increases seemingly the lens parameters already at $B \sim
10^{-5}$ ($ \sim 10^{7}-10^{8} G$).

The Ernst model we used here is certainly is a rough
approximation to a real situation, mainly because magnetic field is
not uniform, but has rather complicated distribution in large scale
space-time. Also, it would be difficult to get a source and an
observer exactly in the equatorial plane. Therefore our estimations
are more representing the illustrative idea that the magnetic fields
create some confinement which increase the bending angle and time
delay, and are not expected to give some high precision results.
We hope, nevertheless, that the influence of magnetic field
upon black hole lensing might be observable in the future.

\section{Conclusion}

We study lensing in the Ernst-Schwarzschild space-time, to
estimate  lens effects of a black hole immersed in a magnetic
field. We showed that already small magnetic field $B \sim
10^{-10}$ (in geometric units), which does not distort the
geometry of the black hole near the event horizon, nevertheless
increases seemingly the bending angle and time delay for a ray of
light propagating in vicinity of a super-massive black hole. This
happens because the magnetic field is supposed to exist even far
from black hole, thereby exerting considerable metric influence on
a ray of light during the whole way from a source to an observer. Yet,
this magnetic field is many order less than the real galactic magnetic
field $B_{g} \sim 10^{-18}$. On
the contrary, estimations in the region close to a black hole,
within the same Ernst-Schwarzschild model, show that strong
magnetic field in the central region near a black hole should give
rise to non-negligible increase in the bending angle and time
delay.

As is known, large-scale magnetic field in Universe has both
poloidal and toroidal components. Yet, it is poloidal component,
which is dominant. Therefore Ernst solution, where the magnetic
field stipulates a single direction in space is reasonable
approximation. Another point is, that very strong magnetic field
in the centre of galaxy, decays with distance. Nevertheless, we
are still able to use the Ernst model for estimations of lens
effects because of the following reason: the lens effect is
calculated by formulas (13) and (14) where integrals in $r$ are
taken from the position of source, through point of minimal
approach, until the position of observer. Therefore if $B$ decays
with $r$, one can consider small interval of $r$, where $B$ can be
considered as approximately homogeneous, and can use the minimal
value of $B$ there. As we have learnt before, the greater $B$
leads to larger bending angle, so that, when we estimated the
bending angle for weak galactic or intergalactic magnetic fields,
we did not take into consideration very strong field in the centre
of galaxy. Thus, our obtained corrections might be less than real.
When investigating the influence of the strong magnetic field in
the centre of galaxy we considered contribution to the bending
angle in a very small region of 20 $r_{g}$ around the centre,
thereby keeping only dominant contribution from the region where
magnetic field is enormously strong. Probably, for our estimations
for the centre of galaxy, one could consider roughly, the magnetic
field inside of the Active Galactic Nuclei be constant. Yet
nowadays, it seems we do not have reliable detailed description of
behavior of magnetic fields in the centre of galaxy. Recent review
of observational data on the behavior of the magnetic field in the
centre of our galaxy can be found for instance in \cite{AGN} and
references therein.

\vspace{4mm}

\begin{acknowledgments}

This work was supported by \emph{Funda\c{c}\~{a}o de Amparo
\`{a} Pesquisa do Estado de S\~{a}o Paulo (FAPESP)}, Brazil.

\end{acknowledgments}

\end{document}